\documentclass[smallextended]{svjour3}       
\smartqed  
\usepackage{graphicx}
\usepackage{amsmath,amsfonts,amssymb,revsymb}
\begin{document}

\title{Strong energy condition and the repulsive character of $f(R)$  gravity}

\author{Crislane S. Santos  \and Janilo Santos \and Salvatore Capozziello \and Jailson S. Alcaniz}


\institute{Crislane S. Santos \at
              Universidade Federal dos Vales do Jequitinhonha e Mucuri, 39100-000 Diamantina - MG, Brazil
               \at
              Departamento de F\'{i}sica Te\'{o}rica e Experimental, Universidade Federal do Rio Grande do Norte, 59072-970 Natal - RN, Brazil \\  \email{crislane@dfte.ufrn.br}           
           \and
           Janilo Santos \at
           Departamento de F\'{i}sica Te\'{o}rica e Experimental, Universidade Federal do Rio Grande do Norte, 59072-970 Natal - RN, Brazil \\
           \email{janilo@dfte.ufrn.br}
           \and
           Salvatore Capozziello \at
              Dipartimento di  Fisica "E. Pancini",  Universit\`{a} di Napoli "Federico II", Naples, Italy
              \at
              Istituto Nazionale di  Fisica Nucleare (INFN) Sez. di Napoli,
              Compl. Univ. di Monte S. Angelo, Edificio G, Via Cinthia, I-80126, Naples, Italy
 \at
Gran Sasso Science Institute,  Viale F. Crispi, 7, I-67100, L'Aquila, Italy
  \at
Tomsk State Pedagogical University, 634061 Tomsk, Russia  634061\\
              \email{capozziello@na.infn.it}
              \and
            Jailson S. Alcaniz  \at
                Observat\'{o}rio Nacional, Rio de Janeiro - RJ, 20921-400 Brazil \\
              \email{alcaniz@on.br}
}

\date{Received: date / Accepted: date}

\maketitle

\begin{abstract}
The Raychaudhuri  equation enables to examine the whole spacetime structure without specific solutions of Einstein's equations, playing  a central role for the understanding of the gravitational interaction in Cosmology. In  General Relativity,   without considering a cosmological constant, a non-positive contribution in the Raychaudhuri equation is usually interpreted as the manifestation of the attractive character of gravity. In this case, particular energy conditions -- indeed the strong energy condition -- must be assumed in order to guarantee the attractive character. In the context of $f(R)$  gravity, however, even assuming the standard energy conditions one may have a positive contribution to the Raychaudhuri equation. Besides providing a simple way to explain the observed cosmic acceleration, this fact opens the possibility of a repulsive character of this kind of gravity. In order to discuss physical bounds on $f(R)$ models, we address the attractive/non-attractive character of $f(R)$ gravity considering the Raychaudhuri equation and assuming  the strong energy condition along with recent estimates of the cosmographic parameters.
\end{abstract}



\section{Introduction}

The observational evidence of the accelerated expansion of the Universe~\cite{obs_data} have been the main reason for a revision of the cosmological evolution as predicted by General Relativity (GR) and the standard model of elementary particles. In principle, this problem can be circumvented either by introducing a term representing a new kind of universal fluid, the so-called dark energy, in the Einstein equations or by considering modifications and extensions of the Einstein theory of gravity. A particular extension of GR, known as $f(R)$ gravity, has received a lot of attention in the last years, mainly from the cosmological viewpoint. The paradigm consists in relaxing the hypothesis that the action of gravitational interaction is strictly linear in the Ricci scalar $R$, as in the case of the Hilbert-Einstein action, and assuming a general function of $R$ that can be fixed by observations and experiments.

The cosmological interest in $f(R)$ gravity comes from the fact that these theories  naturally exhibit a late accelerating expansion of the Universe without need of exotic matter fields such as dark energy (for reviews on this subject, see Ref.~\cite{reviews}). Much effort has been expended so far in order to limit the freedom of different  functional forms  which are possible for $f(R)$ models. Recently, observational constraints from several cosmological data sets have been explored~\cite{cosmological_tests}. General principles such as nonlocal causal structure~\cite{causal-structure}, energy conditions~\cite{EC-I,EC-II}, have also been taken into account in order to restrict the solution space and to clarify some of  subtleties related to $f(R)$ gravity.

An important aspect worth emphasizing when discussing this issue is that the energy conditions were initially formulated by Hawking and Ellis in the context of GR~\cite{Hawking}. As already noted by the authors of Ref. \cite{EC-I}, $f(R)$ models are described in the so-called Jordan frame while Einstein's gravity is formulated in the Einstein frame. However, it can be shown that any $f(R)$ theory is mathematically equivalent, via conformal transformation, to Einstein's gravity with a minimally coupled scalar field~\cite{conf_equivalence-I}. Perhaps guided by these studies, some authors have translated the energy conditions directly from GR, imposing them on an effective pressure and effective energy density defined by the related effective energy-momentum tensor~\cite{bergliaffa}.  In order to test the viability of such a procedure, the authors of Ref. \cite{EC-I} generalized the energy conditions for $f(R)$ gravity which has been followed by many authors (see, e.g., Ref. \cite{EC-II}) and generalized to other modified theories of gravity~\cite{EC-modified_gravity}. Other approaches, however, consider the new terms appearing in the equations of motion must be understood as possessing geometrical meaning only~\cite{lambiase,JCAP-1,capozziello,mimoso}. Indeed, the physical connection between conformal frames, or, in other words, the problem whether the physical information\footnote{The energy conditions contain physical information.} contained in the theory is preserved under conformal transformations, is still a contentious issue~\cite{conf_equivalence-II}.

In this paper, following the investigation of Ref.~\cite{EC-I} and the developments presented in Ref.~\cite{JCAP-1}, we consider the strong energy condition (SEC) in the framework of $f(R)$ gravity and use recent estimated values of deceleration, jerk and snap cosmographic parameters of Friedmann-Robertson-Walker (FRW) flat geometry to impose bounds on the parameters of a paradigmatic class of $f(R)$ gravity models. We also address the attractive/non-attractive character of $f(R)$ gravity by deriving explicit bounds on some physical quantities 
in a flat FRW geometry.

The paper is organized as follows. Sec. II discuss the Raychaudhuri equation where, in particular, we point out the role of the curvature in order to obtain focusing or defocusing bundles of geodesics.  $f(R)$ gravity and strong energy condition are presented in Sec. III. Specifically, we explicit  the relation between the mean Gaussian curvature of geodesic surfaces, the Raychaudhuri equation and the field equations of $f(R)$ gravity. This relation allows to set conditions for acceleration/deceleration for the given models. Sec. IV consists of a cosmographic analysis in FRW geometry.  The observed values of cosmographic parameters allows to fix $f(R) $ models in view of accelerated/decelerated expansion. Discussion and conclusions are presented in Sec. V. We use a metric signature $(-,+,+,+)$ and our definition of the Riemann tensor is  $ R^{\rho}_{\sigma\mu\nu}\equiv\partial_{\mu}\Gamma^{\rho}_{\nu\sigma}- \ldots $,
$R_{\mu\nu}\equiv R_{\mu\lambda\nu}^{\lambda}$ defines the Ricci tensor and $R=R^{\mu}_{\mu}$ is the Ricci scalar.

\section{The Raychaudhuri Equation}

When examining the whole  spacetime structure without specific solutions of Einstein's equations, the Raychaudhuri equation~\cite{Raychaudhuri} plays a central role in the understanding of the gravitational attraction. Such an evolution equation describes the geodesic motion of nearby particles without making assumptions about homogeneity and  isotropy of the spacetime. In GR,  without a cosmological constant, particular energy conditions -- indeed the strong energy condition -- has to be assumed in order to guarantee the attractive character of the theory~\cite{Carroll-Book}. Positive contributions coming from  the spacetime geometry to the Raychaudhuri equation are usually interpreted as violation of the SEC or the null energy conditions requirements~\cite{Poisson}. In the context of $f(R)$ gravity, however, we may have a positive contribution to Raychaudhuri equation even assuming these standard energy conditions, which gives the possibility of a repulsive character of this kind of gravity (we will discuss this point later). The Raychaudhuri equation for a congruence of timelike geodesics and its tangent vector field $\xi^{\mu}=dx^{\mu}/d\tau$ is written as
\begin{equation}
\frac{d\theta}{d\tau}= - \frac{\theta^2}{3}-\sigma_{\mu\nu}\sigma^{\mu\nu}+w_{\mu\nu}w^{\mu\nu}
-R_{\mu\nu}\xi^{\mu}\xi^{\nu},
\label{Raychaudhuri tempo}
\end{equation}
where $\theta,\sigma^{\mu\nu},w^{\mu\nu}$ are respectively the expansion, shear and twist of the congruence of geodesics, and $\tau$ is the proper time of an observer moving along a geodesic. It is worth stressing that (\ref{Raychaudhuri tempo}) gives the expansion rate of the congruences as seen by comoving observers and, due to its spatial character, the quantities $\sigma_{\mu\nu}\sigma^{\mu\nu}$ and $w_{\mu\nu}w^{\mu\nu}$ are non-negative scalars over a given spatial section. For the contribution of the term $R_{\mu\nu}\xi^{\mu}\xi^{\nu}$,  there appear three possibilities: negative, positive or zero contribution, depending both on the point of the manifold and on the direction of the vector $\xi^{\mu}$ at that point. However, it can be shown that the expression $R_{\mu\nu}\xi^{\mu}\xi^{\nu}$ is related to the Gaussian curvature $K_{(0A)}$ of the geodesic surface generated by $\xi^{\mu}_{(0)}$ and $\xi^{\mu}_{(A)}$ \cite{eisenhart}. The sum of all the Gaussian curvatures of the geodesic surfaces is
$\sum_{A=1}^{3}K_{(0A)} = - R_{\mu\nu}\xi^{\mu}_{(0)}\xi^{\nu}_{(0)}$, which is called ``the mean curvature of the space" in the direction $\xi^{\mu}_{(0)}$ by  Eisenhart in Ref.~\cite{eisenhart}. Here, we denote it simply as ${M}_{\xi^{\mu}}\equiv - R_{\mu\nu}\xi^{\mu}\xi^{\nu}$ (see also Ref. \cite{JCAP-2} for more details).

Although the Eq. (\ref{Raychaudhuri tempo}) has only geometrical meaning, once one chooses a particular theory of gravitation, its contribution, via the field equations of motion to the kinematic of the congruences, is carried out through the terms $-R_{\mu\nu}\xi^{\mu}\xi^{\nu}$.
A negative contribution to Raychaudhury equation (focusing, $d\theta/d\tau < 0$) is usually interpreted as manifestation of the attractive character of the theory of gravity. The contribution of ${M}_{\xi^{\mu}}$  has then a clear geometrical interpretation~\cite{JCAP-1,JCAP-2} which is resumed as:
\begin{itemize}
  \item[(i)] ${M}_{\xi^{\mu}}>0 \Rightarrow$ Positive contribution (a condition necessary but not sufficient to geodesic defocusing);
  \item[(ii)] ${M}_{\xi^{\mu}}<0 \Rightarrow$ Negative contribution (geodesic focusing);
  \item[(iii)] ${M}_{\xi^{\mu}}=0 \Rightarrow$ Zero contribution.
\end{itemize}

In what follows we consider congruence of geodesics for which $w_{\mu\nu}=0$\footnote{This always holds if the tangent vector field is locally hypersurface orthogonal.} and rewrite equation (\ref{Raychaudhuri tempo}) as
\begin{equation}
\frac{d\theta}{d\tau}= - \frac{\theta^2}{3}-\sigma_{\mu\nu}\sigma^{\mu\nu} + M_{\xi^{\mu}}\;.
\label{Raychaudhuri tempoII}
\end{equation}
From this formula it is clear the role of ${M}_{\xi^{\mu}}$, in Raychaudhuri equation,  for the focusing or defocusing  the congruence of geodesics. A positive contribution from $M$ is a necessary condition to get accelerated expansion and, also if it is  not a sufficient condition, it may lead to non-attractive gravity.  In fact, as commented in Ref.~\cite{Visser}, the Raychaudhuri equation can either use  the congruences to provide information on the Ricci tensor, and hence on the stress-energy tensor, via the equations of motion, or can use  the stress-energy tensor to provide information about the congruences. As the Ricci tensor is in general unknown from the beginning\footnote{Specially in $f(R)$ theory of gravity.}, we take a step forward by using the gravitational field equations and imposing the SEC on the stress-energy tensor in order to provide information about the congruences. We intend to test the limit of attractiveness/non-attractiveness of $f(R)$ gravity adopting the SEC as a physical paradigm.

\section{$f(R)$ gravity and the strong energy condition}

The action that defines an $f(R)$ gravity is given by
\begin{equation}
\label{actionJF}
S = \frac{1}{2\kappa^2}\int d^4x\sqrt{-g}[R + f(R)] + S_m\,,
\end{equation}
where $\kappa^2=8\pi G$, $g$ is the determinant of the metric tensor and $S_m$ is the standard action for the matter
fields. Varying  the action~(\ref{actionJF}) with respect to the metric,  we obtain the field equations
\begin{equation}
\label{einstein-modified}
(1+f')R_{\mu\nu}-\frac{1}{2}(R+f) g_{\mu\nu}
-(\nabla_{\mu}\nabla_{\nu}-g_{\mu\nu}\nabla^{\alpha}\nabla_{\alpha})f'=8\pi G T_{\mu\nu},
\end{equation}
where $f'=df/dR$, 
and the stress-energy tensor is defined as
\begin{equation}\label{energy-momentum}
T_{\mu\nu}= \frac{-2}{\sqrt{-g}}\,\frac{\delta S_{m}}{\delta g^{\mu\nu}}\,.
\end{equation}
Contracting the equation (\ref{einstein-modified}) with $\xi^{\mu}\xi^{\nu}$, where $\xi^{\mu}$ is a normalized timelike vector ($\xi^{\mu}\xi_{\mu}=-1$), and taking into account its trace, we obtain, according to our definition for ${M}_{\xi^{\mu}}$, the expression
\begin{equation}
M_{\xi^{\mu}}=\frac{Rf' -f + (\nabla^{\alpha}\nabla_{\alpha} - 2\xi^{\mu}\xi^{\nu}\nabla_{\mu}\nabla_{\nu})f'
   - 16\pi G(T_{\mu\nu}\xi^{\mu}\xi^{\nu}+T/2)}{2(1+f')},
\label{secf2}
\end{equation}
where $T$ is the trace of the stress-energy tensor.
The SEC and the null energy condition (NEC) state that~\cite{Hawking}
\begin{itemize}
\item[SEC:] $\qquad T_{\mu\nu}\xi^{\mu}\xi^{\nu}+ T/2 \geq 0$;
\item[NEC] $ \qquad T_{\mu\nu}\kappa^{\mu}\kappa^{\nu}\geq 0$,
\end{itemize}
where $\kappa^{\mu}$ is a null vector.
For a perfect fluid, characterized by a density $\rho$ and a pressure $p$, the SEC states that we must have, besides $\rho + p \geq 0$, the sum $\rho + 3p\geq 0$, while the NEC implies that $\rho + p\geq 0$~\cite{Carroll-Book}. In GR the observed accelerated expansion of the Universe implies violation of the SEC. Since the requirement of NEC is also contained in the SEC, in what follows we consider in our study only the SEC restrictions. Imposing the SEC to the standard cosmological fluids in the expression (\ref{secf2}) we obtain
\begin{equation}
M_{\xi^{\mu}}\leq\frac{ Rf'-f+(\nabla^{\alpha}\nabla_{\alpha} -2\xi^{\mu}\xi^{\nu}\nabla_{\mu}\nabla_{\nu})f'}{2(1+f')},
\label{mean curvature time2}
\end{equation}
which we interpret as an upper bound to the contribution of spacetime geometry (gravity) to the Raychaudhuri equation related to timelike geodesics. Observe that, for $f(R)=-2\Lambda$, which gives the Einstein equations of GR with a cosmological constant $\Lambda$ (see Eq. (\ref{einstein-modified})), we get from (\ref{mean curvature time2}) that $M_{\xi^{\mu}}\leq\Lambda$.
Hence we have that, for timelike geodesics, a positive contribution to the Raychaudhuri equation from spacetime geometry ($M_{\xi^{\mu}}\geq0$) is possible provided $\Lambda>0$. It corresponds to the correct sign of $\Lambda$ which provides cosmic acceleration in the standard $\Lambda$CDM model. Thus, if one considers GR  as a  $f(R)=-2\Lambda$ theory of gravity, accelerated cosmic expansion is possible without violating the SEC.  For a more general $f(R)$ gravity, even assuming the validity of the SEC, the sign of $M_{\xi^{\mu}}$ still remains undetermined, and one has the possibility of a non-attractive gravity in these theories. We shall explore this possibility in the next section and look for cosmographic constraints in the geometrical background of a flat FRW metric.

\section{Cosmographic Constraints}

Equation (\ref{mean curvature time2}) is an upper bound to the contribution of spacetime geometry obtained by imposing the SEC inequality. Here we use this inequality, together with known values of the so called cosmographic parameters, in order to examine the attractive/non-attractive character of a given $f(R)$ model in the context of FRW flat geometry. In \cite{JCAP-2},  the authors considered a general congruence of timelike geodesics and examined the conditions to have $M_{\xi} >0$. Here we consider the fundamental congruence of geodesics in FRW as the one where the observer is comoving, that is, $\xi^{\mu}_{FRW}=\partial_t\,$. In this case the equation (\ref{mean curvature time2}) can be written as
\begin{eqnarray}  \label{mean curvature time3}
M_{\xi^{\mu}_{FRW}}\leq -\, \frac{f - Rf' + 3(\ddot{R}+H\dot{R})f'' +3\dot{R}^{2}f'''}{2(1+f')}\,.
\end{eqnarray}
We want to investigate such a bound in terms of the cosmographic parameters, i.e. the Hubble $H$, deceleration $q$, jerk $j$, and snap $s$ parameters, defined respectively as
\begin{eqnarray}
H=\frac{\dot{a}}{a},\quad q=-\frac{1}{H^{2}}\frac{\ddot{a}}{a}, \quad j= \frac{1}{H^{3}}\frac{\dddot{a}}{a},\quad s=\frac{1}{H^{4}}\frac{\ddddot{a}}{a}\,.
\label{parametros}
\end{eqnarray}
Firstly we have to  express the Ricci scalar and its derivatives in terms of these parameters:
\begin{eqnarray}
&&R = 6H^{2}(1-q),\nonumber\\
&&\dot{R} = 6H^{3}(j-q-2),\nonumber\\
&&\ddot{R} =  6H^{4}(s+q^{2}+8q+6)\,.
\label{ricci scalar jerk}
\end{eqnarray}
After some calculations, and using the relations (\ref{ricci scalar jerk}), we can write $M_{\xi^{\mu}_{FRW}}$ in terms of the cosmographic parameters for a general form of $f(R)$:
\begin{equation}
M_{\xi^{\mu}_{FRW}}\leq\frac{- f/2 + c_1f' + c_2f'' + c_3f'''}{1+f'},
\label{geral4time}
\end{equation}
where the coefficients $c_i$ are given in terms of the cosmographic parameters as
\begin{eqnarray}
c_1 & = & 3(1-q)H^{2} \nonumber \\
c_2 & = & -9(s+j + q^{2}+7q+4)H^{4} \nonumber \\
c_3 & = & - 54(j-q-2)^{2}H^{6}\,.
\end{eqnarray}
Inequality (\ref{geral4time}) is valid at any time, to the extent that the SEC is also valid at any time.

In the case of a FRW geometry,  we have $\sigma_{\mu\nu}=0$ so that the Raychaudhuri equation reduces to
\begin{equation}
\frac{d\theta}{d\tau}= - \frac{\theta^2}{3} + M_{\xi^{\mu}_{FRW}}.
\label{Raychaudhuri_RW1}
\end{equation}
Note that  GR implies  $f(R)=0$ and, from Eq.~(\ref{secf2}), we obtain $M_{\xi^{\mu}}=-4\pi G(\rho + 3p)$ for a perfect fluid considered as the matter source. Thus, in GR with a FRW geometry ($3\ddot{a}/a = - 4\pi G[\rho + 3p]$), the Raychaudhuri equation gives $d\theta/d\tau = - 3H^2(1+q)$, showing that geodesic defocusing only appears if $q< -1$. However, for theories of gravity more general  than GR, we can have, from (\ref{Raychaudhuri_RW1}), that  $M_{\xi^{\mu}_{FRW}}\geq \theta^2/3$, turning $d\theta/d\tau$ positive even for $q> -1$. This contributes to a non-attractive character of gravity in these theories, and also affect the proof of the celebrated singularity theorems due to Penrose, Hawking and Geroch\footnote{As well known, having $d\theta/d\tau <0$ in Raychaudhuri equation play a key role in the proof of some singularity theorems~\cite{Raychaudhuri}.}.

While the action (\ref{actionJF}) may not represent the final  theory of modified gravity, it could contain the information necessary to act as an effective field theory capable of describing correctly the phenomenology of gravitation. In this case, a particular $f(R)$ model, seen as an effective field theory, may have a limited region of applicability. This means to lose general prescription and to obtain results valid only to the considered $f(R)$ function.  Alternatively, one could restrict to analytic $f(R)$ functions so that it can be expanded about a certain $R=R_0$ as a power series
\begin{equation} \label{power-series}
f(R) = \sum_{n=-\infty}^{+\infty}\,a_n(R-R_0)^n.
\end{equation}
Several models of cosmological interest can be expressed as (\ref{power-series}). However, as a case to study, in order to explicit our calculations and understand the contribution of spacetime geometry according to the free parameters of a given $f(R)$ theory of gravitation,
in what follows we will use the inequality (\ref{geral4time}) to examine the behavior of the paradigmatic class of $f(R)$ models given by
\begin{equation} \label{paradig}
f(R) = \frac{\alpha}{R^n}\,.
\end{equation}
This class of $f(R)$ gravity encompass a wide variety of proposals in the scientific literature. For instance, Starobinsky considered in \cite{starobinsky} the case with $n =-2$ as a viable scenario of primordial inflation. More recently, the same case was considered in \cite{cembranos} as a possibility of explaining dark matter.  The case $n=1$, specially for $\alpha <0$,  was presented in \cite{capozziello1,carroll} as a possible mechanism to provide cosmological acceleration; although it is currently ruled out due to the so called Dolgov-Kawasaki instability~\cite{dolgov}. In~\cite{faraoni},  it was proved that these theories also suffer from the Dolgov-Kawasaki instability for negative values of $\alpha$ and $n>0$ not restricted to be an integer. In principle, negative and positive powers can contribute to cosmic dynamics as discussed in \cite{orly}. In what follows,  we consider the cases  $\alpha$ negative\footnote{$\alpha <0$ is fundamental to reproduce the $\Lambda$CDM model, as will be seen bellow.} and $n$ a real number. We examine the behaviour of this class of models in what concerns the magnitude of $\alpha$ and $n$ and the cosmographic parameters.

Taking into account the first of the relations (\ref{ricci scalar jerk}),  we can derive $\alpha$ in (\ref{paradig}) as $\alpha = f_0[6H_0^2(1-q_0)]^n$ where $f_0 = f(R_0)$ and $R_0$, $H_0$ are, respectively, the Ricci scalar curvature and the Hubble parameter for the present time. Note that for $n=0$ we obtain $\alpha = f_0$, thus if we take $f_0=-2\Lambda$ ($\Lambda$ a positive constant), the quantity ${M}_{\xi^{\mu}}$ for Einstein's gravity with cosmological constant $\Lambda$ is recovered (see (\ref{mean curvature time2}) for $f=-2\Lambda$). We can take a step forward and define the parameter $\Omega_{\Lambda}\equiv \Lambda/3H_0^2$ such that now $\alpha = -6\Omega_{\Lambda}H_0^{2n+2}[6(1-q_0)]^n$.
Using this relation for $\alpha$ and calculating the derivatives of (\ref{paradig}), we rewrite (\ref{geral4time}) in terms of the cosmographic parameters as a dimensionless inequality
\begin{equation} \label{mean-curvature_time}
\frac{M_{\xi^{\mu}_{FRW}}}{H^2} \leq \frac{6\Omega_{\Lambda} Q(n)}{\frac{n\Omega_{\Lambda}}{1-q} +
\left( \frac{1-q}{1-q_0} \right)^{n}\left( \frac{H}{H_0} \right)^{2n+2}}\;,
\end{equation}
where $Q(n)$ is a third degree polynomial in $n$
\begin{equation} \label{poly}
Q(n)=(n+1)\left(A(t)n^2 + B(t)n + \frac{1}{2}\right)\,,
\end{equation}
with the coefficients given in terms of the cosmographic parameters as
\begin{equation} \label{A_coeff}
A(t)= -\frac{(j-q -2)^2}{4(1-q)^3}\,,
\end{equation}
\begin{equation}  \label{B_coeff}
B(t)=2A(t) + \frac{q^2 + 7q + 4 + j + s}{4(1-q)^2}\,.
\end{equation}
Note that for $n=0$, the relation  (\ref{mean-curvature_time}) gives $M_{\xi^{\mu}_{FRW}}\leq \Lambda$, as expected for  Einstein's gravity with a cosmological constant, while for $n=-1$ (Einstein's gravity with a modified coupling constant) one obtain (under the assumption that $\Omega_{\Lambda}\neq 1- q_0$)
$M_{\xi^{\mu}_{FRW}}=0$ at any time.

It is worth  noticing  that inequality (\ref{mean-curvature_time}) was derived  assuming  that the SEC holds over all  the cosmological eras and that the theory of gravity can be described by action (\ref{actionJF}) with $f(R)$ given by (\ref{paradig}). As such, it
can be used to impose bounds on the free parameters of the class (\ref{paradig}) of $f(R)$ models  based only on the SEC and
present day ($t=t_0$) estimated values of cosmographic parameters.
In this case the inequality (\ref{mean-curvature_time}) reduces to
\begin{equation} \label{mean-curvature_time-II}
\frac{M_{\xi^{\mu}_{FRW}}}{H_0^2} \leq \frac{6\Omega_{\Lambda}}{\frac{n\,\Omega_{\Lambda}}{1-q_0}+1 }  \,Q(n) \,,
\end{equation}
where the coefficients of $Q(n)$ are given by (\ref{A_coeff})--(\ref{B_coeff}) taken at $t=t_0$.
Inequality (\ref{mean-curvature_time-II}) provides an upper limit on the contribution ${M}_{\xi^{\mu}}\equiv - R_{\mu\nu}\xi^{\mu}\xi^{\nu}$ imposed by the SEC on $f(R)=\alpha/R^n$ theories of gravity for parameters $\{n,\alpha\}$ based on the present day estimated values of the cosmographic parameters. Positive values for $M_{\xi^{\mu}}$, eventually surpassing the negative values in Eq. (\ref{Raychaudhuri_RW1}) leading to $d\theta/d\tau >0$ (geodesic defocusing), may be interpreted as a repulsive force~\cite{JCAP-1}. Therefore, depending on the values of the parameters in (\ref{paradig}), we may have $M_{\xi^{\mu}_{FRW}}>\theta^2/3$. Taking into account that $\theta = 3H$ in the flat FRW geometry, we have a lower limit which, in terms of the present-day Hubble parameter, reads $M_{\xi^{\mu}_{FRW}}/H_0^2 >3$. In other words, these $f(R)$ models, for which the parameters are such that the value of ${M}_{\xi^{\mu}}$ in units of $H_0^2$ are greater than 3, give geodesic defocusing and can be interpreted as a repulsive gravity.

\begin{figure}
  \includegraphics[width=1.0\textwidth]{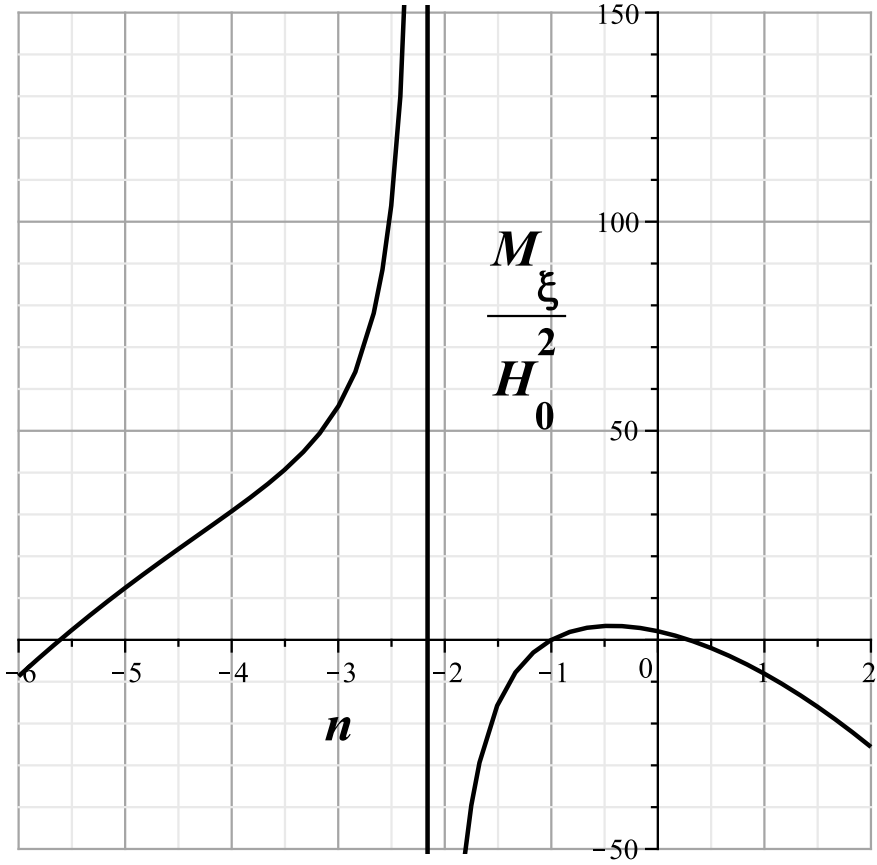}
\caption{Behavior of $M_{\xi^{\mu}}/H_0^2$  (Eq. \ref{mean-curvature_time-II}) for the mean values of the cosmographic parameters specified in the text. We have assumed $\Omega_{\Lambda}=0.69$. }
\label{M_mean}       
\end{figure}

In order to have a better understanding of this model, let us  examine the behavior of the inequality (\ref{mean-curvature_time-II}) taking into account the constraints coming from  the cosmographic parameters determined in Ref.~\cite{cosmography}. We use their mean values estimated from data combinations of SNeIa + GRB + BAO + CMB (see Table I in \cite{cosmography}), which are
$S_0=\{q_0=-0.49\pm 0.29,j_0=-0.50\pm 4.74,s_0=-9.31\pm 42.96\}$. In Fig.\ref{M_mean},  we show a plot of (\ref{mean-curvature_time-II}) where we have assumed $\Omega_{\Lambda}=0.69$ from~\cite{planck_XIII}.

Observing this plot, we find that the attractive/repulsive character of this class of $f(R)$ gravity depends smoothly on the free parameter $n$ for the given set $S_0$. However, for $n=-(1-q_0)/\Omega_{\Lambda}\approx -2.16$ (vertical line in Fig.\ref{M_mean}) we have a strong singularity, where gravity abruptly changes from extremely attractive to extremely repulsive. Even taking into account that this critical $n$ depends on the values of $q_0$ and $\Omega_{\Lambda}$, this is interestingly very close to the $f(R)\propto R^2$ Starobinsky inflation~\cite{starobinsky}. However, the present study makes use of the estimated values of the set $S_0$, which are valid only for redshifts $z<1$. We have also examined the dependence of $M_{\xi^{\mu}_{FRW}}/{H_0^2}$ with respect to $\Omega_{\Lambda}$ and found that it is not sensitive to small changes of this parameter. On the overall we find that for the upper values of the set $S_0$ we have $d\theta/d\tau > 0$, even obeying the SEC, when $0.04<n<6.18$. This change in the behavior of the expansion $\theta$ is indicative that this interval for $n$ should be discarded for this class of $f(R)$.

\begin{figure}
  \includegraphics[width=1.0\textwidth]{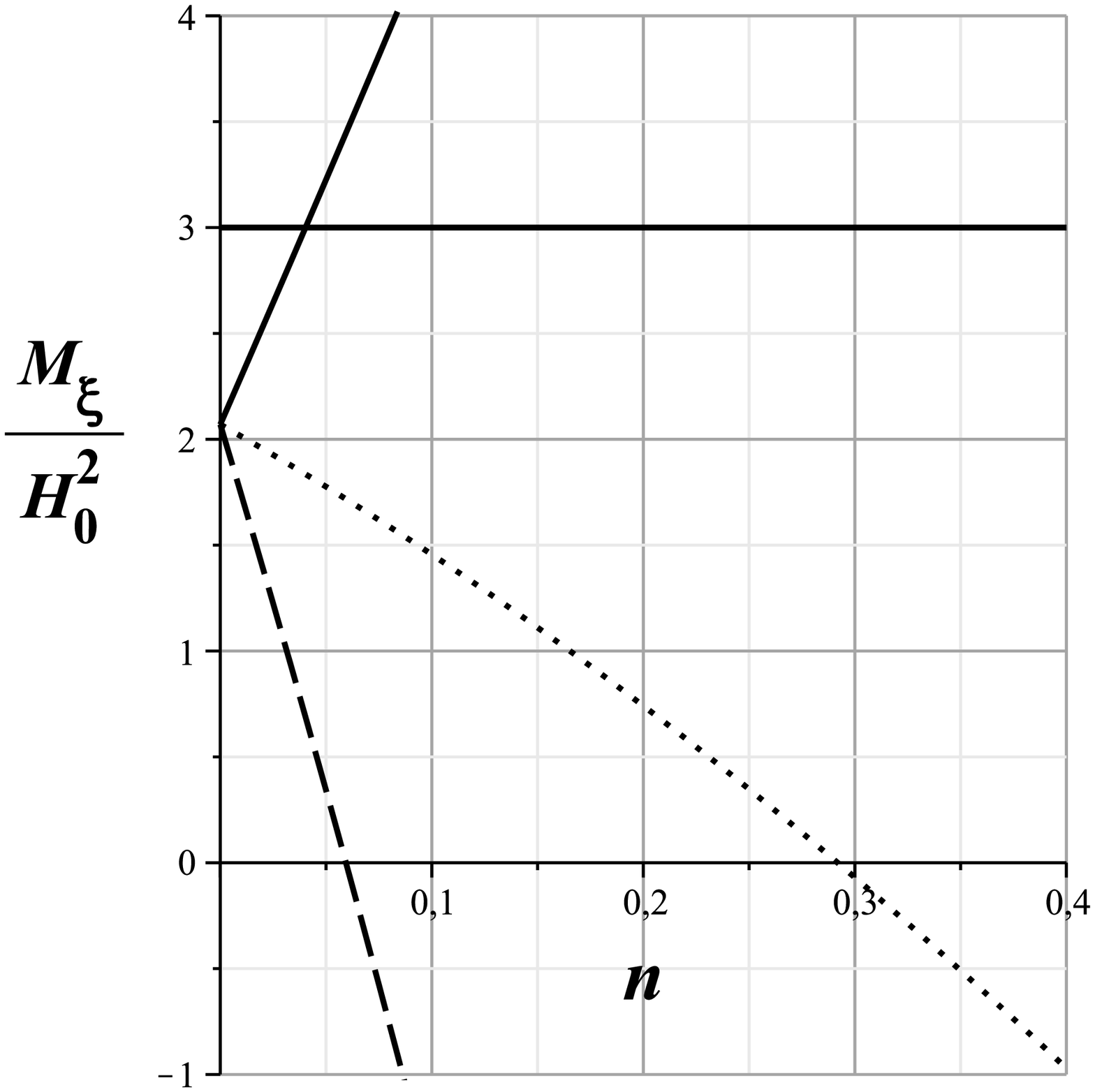}
\caption{Behavior of $M_{\xi^{\mu}}/H_0^2$  (Eq. \ref{mean-curvature_time-II}) in the neighborhood of $n\geq 0$ for the lower (dashed line), mean (doted line) and upper (solid line) values of the cosmographic parameters specified in the text. The horizontal line is $M_{\xi^{\mu}}/H_0^2=3$ and we have taken $\Omega_{\Lambda} =$ 0.69. }
\label{M_viz}    
\end{figure}

Plots of (\ref{mean-curvature_time-II}), in the neighborhood of $n\geq 0$,  are shown in Fig. \ref{M_viz} for the lower, mean and upper values of the cosmographic parameters in the set $S_0$. Examining these curves we find that contribution to $M_{\xi^{\mu}}$ positive ($0<M_{\xi^{\mu}}/{H_0^2}<3$) happens for: (i) $0<n<0.06$,  if the cosmographic parameters are given by the lower values of $S_0$ (dashed line),  (ii) $0<n<0.29$, if we take the mean values of $S_0$ (doted line) and (iii), if the upper values of $S_0$ are used, $0<n<0.04$ (solid line) or $6.18<n<6.26$ (this second limits for $n$ are not shown in Fig. \ref{M_viz}). Taking into account that $n=0$ ($\Lambda$CDM model) gives $M_{\xi^{\mu}}/H_0^2 =3\Omega_{\Lambda}=2.07$, we find that upper values of $S_0$ provide faster  expansion for any $0<n<6.20$ while the lower and mean values of $S_0$ tend to slow down the expansion in comparison with
$\Lambda$CDM model.

In Ref.~\cite{carroll} the authors present an $f(R)$ gravity, which we name CDTT model, given by $f(R)=-\mu^4/R$.  In order to compare the CDTT model with our study, note that we have taken $f_0/H_0^2 = -6\Omega_{\Lambda}$, so the parameters $\Omega_{\Lambda}$ and $\mu$ are related by $\Omega_{\Lambda}=(\mu/H_0)^4/[36(1-q_0)]$. The authors in \cite{carroll} claim that choosing $\mu\approx H_0$ their model could, in principle, explain the present accelerated expansion of the Universe without need of dark energy. In our study this amounts to choose our parameter as $\Omega_{\Lambda}=1/[36(1-q_0)]$. However, for the mean value in $S_0$ ($q_0=-0.49$), this corresponds to take $\Omega_{\Lambda}=0.02$ what gives  $M_{\xi}(n=1)/{H_0^2}=-0.32$. Negative values for the present $M_{\xi^{\mu}_{FRW}}$ contribution means that the Universe is not accelerating. In other words, if we demand {\it ab initio} that the SEC is valid, the CDTT model~\cite{carroll} cannot give rise to an  accelerated expansion at the present era.
On the other hand, the instability in the Ricci curvature scalar, pointed out by Dolgov and Kawasaki~\cite{dolgov}, and further generalized by Faraoni~\cite{faraoni} for this kind of gravity, does not appear in $M$ since, as we have seen above, $M_{\xi}(n=1)/{H_0^2}=-0.32$ is not exactly a problem. However, taking $f_0/H_0^2=-6\Omega_{\Lambda}=-4.14$ as we did in our analysis, one may have $M_{\xi}(n=1)/{H_0^2}\ll 0$ or $M_{\xi}(n=1)/{H_0^2}\gg 3$, providing  strong attraction or strong repulsion, respectively, depending of the values of  cosmographic parameters.
For more general CDTT models given by $f(R)=-\mu^{2(n+1)}/R^n$, our parameter $\Omega_{\Lambda}$ relates to the $\mu$ parameter through $6\Omega_{\Lambda}=(\mu/H_0)^{2(n+1)}/[6(1-q_0)]^n$, such that, for $\mu\approx H_0$, it provides $6\Omega_{\Lambda}=1/[6(1-q_0)]^n$.

Finally  some comments about constraining the sign of $M_{\xi^{\mu}_{FRW}}$ as given by equations (16)--(19) are necessary.
Although we  have not the values for the cosmographic parameters over time, if we make the reasonable assumption that $q(t)< 1$, we see that the coefficient $A(t)$, given by (\ref{A_coeff}), is always negative. The three roots of $Q(n)$ are $\{-1,n_{\pm}\}$ with
\begin{equation}\label{Q-roots}
n_{\pm} = \frac{B(t)}{2|A(t)|}\left( 1 \pm \sqrt{1+\frac{2|A(t)|}{B^2(t)}} \right),
\end{equation}
where we have taken into account the sign of $A(t)$. Let us note that the product of this roots gives $n_{+}\times n_{-}=-1/|2A(t)|$, so, one root is positive and the other is negative. From (\ref{Q-roots}) we also see that the signals of the roots $n_{\pm}$ are controlled by the sign of the coefficient $B(t)$: if $B(t)>0$ we have $n_{+}>0$ and $n_{-}<0$, the opposite occurs if $B(t)<0$\footnote{In the above analysis for $t=t_0$ we find that $B(t_0)<0$ for the lower and the mean values of the present day cosmographic parameters while $B(t_0)$ is positive for the upper values.}. In summary, we always have only one positive root whatsoever is the sign of $B(t)$. In addition, if we restrict ourselves to $n\geq 0$ in (\ref{paradig}), the sign of $M_{\xi^{\mu}_{FRW}}$ in (\ref{mean-curvature_time})
will be controlled by the second degree polynomial $A(t)n^2 + B(t)n + 1/2$ whose graph is a parabola with the concavity facing down. This means that
$M_{\xi^{\mu}_{FRW}}$ will be negative for $n$ greater than the positive root, thus making accelerated expansion impossible, and $M_{\xi^{\mu}_{FRW}}>0$ for intermediary values $n\geq 0$ yet below the positive root of $Q(n)$. This is what is explicitly shown in Fig. \ref{M_viz} for the lower and mean values of the cosmographic parameters determined in Ref.~\cite{cosmography}.

\section{Final Remarks}

Energy conditions play a fundamental role in setting physical constraints for relativistic theories. In general, the compatibility of source fluids with causality and geodesic structure is determined by the consistency of energy conditions. Considering modified gravity means, in some sense, introducing further geometrical components (e.g. $f(R)$ gravity) and scalar fields (e.g. Brans-Dicke gravity)  that can alter the meaning of energy conditions. However, these further degrees of freedom can be recast as additional effective fluids and then inserted in the  energy conditions. In such a case, accurate considerations have to be developed in order to establish the physical role of fields in  the Jordan frame and in the Einstein frame (see  \cite{capozziello,mimoso} for a detailed discussion).

In this paper, we demonstrated that energy conditions, in particular the SEC, play an important role in selecting repulsive/attractive gravity in  the framework of $f(R)$ gravity. For a given class of $f(R)$ models, we have shown that SEC and Raychaudhuri equation can be combined with cosmographic parameters and then confronted with observations. From a methodological point of view, results indicate that  such an {\it Energy Condition Cosmographic Approach}  can be extremely useful in order to fix viable models. Here we have taken into account only $f(R)$ power-law models.  We have shown that  observations, combined with  SEC, fix the range of  viable powers for selecting  attractive/repulsive  models or, according to dark energy paradigm, accelerated/decelerated models. The method  seems promising in view of further applications to  more realistic physical models. This goal will be the topic of forthcoming studies.

\section*{Acknowledgments}
CSS thanks support by REUNI/CAPES. SC and JSA acknowledge the Program Science Without Borders - CNPq (Brazil), CNPq/400471/2014-0 {\it Theoretical and Observational Aspects of Modified Gravity Theories}. SC also thanks the  COST Action (CANTATA/CA15117), supported by COST (European Cooperation in Science and Technology). JSA is also supported by FAPERJ.


 \end{document}